\documentclass[conference,a4paper]{APSIPA2021}
\usepackage{amsmath}
\usepackage{graphicx}
\usepackage{multirow}
\usepackage{booktabs} 
\usepackage{threeparttable}
\usepackage[backend=biber,style=ieee,]{biblatex}
\usepackage{array}

\usepackage{geometry}
\usepackage{fancyhdr}

\fancypagestyle{firststyle}{
  \fancyhf{}
  \fancyhead[C]{2024 Asia Pacific Signal and Information Processing Association Annual Summit and Conference (APSIPA ASC)}
}

\begin{document}

\title{Improved Architecture for High-resolution Piano Transcription to Efficiently Capture Acoustic Characteristics of Music Signals}

\author{
\authorblockN{
Jinyi Mi,
Sehun Kim and
Tomoki Toda
}

\authorblockA{
% \authorrefmark{1}
Nagoya University, Japan \\
E-mail: \{mi.jinyi, kim.sehun\}@g.sp.m.is.nagoya-u.ac.jp, tomoki@icts.nagoya-u.ac.jp}

% \authorblockA{
% \authorrefmark{2}
% University of Macau, Macau \\
% E-mail: yy@um.edu.mo  Tel/Fax: +853-XXXXXXXX}
}

\maketitle
\thispagestyle{firststyle}
\pagestyle{fancy}

\begin{abstract}
Automatic music transcription (AMT), aiming to convert musical signals into musical notation, is one of the important tasks in music information retrieval. Recently, previous works have applied high-resolution labels, i.e., the continuous onset and offset times of piano notes, as training targets, achieving substantial improvements in transcription performance. However, there still remain some issues to be addressed, e.g., the harmonics of notes are sometimes recognized as false positive notes, and the size of AMT model tends to be larger to improve the transcription performance. To address these issues, we propose an improved high-resolution piano transcription model to well capture specific acoustic characteristics of music signals. First, we employ the Constant-Q Transform as the input representation to better adapt to musical signals. Moreover, we have designed two architectures: the first is based on a convolutional recurrent neural network (CRNN) with dilated convolution, and the second is an encoder-decoder architecture that combines CRNN with a non-autoregressive Transformer decoder. We conduct systematic experiments for our models. Compared to the high-resolution AMT system used as a baseline, our models effectively achieve 1) consistent improvement in note-level metrics, and 2) the significant smaller model size, which shed lights on future work.
\end{abstract}

\section{Introduction}
Automatic music transcription (AMT) has gained considerable research interest in the fields of music signal processing and music information retrieval (MIR) for several decades \cite{benetos2018_Review}. The object of AMT is to convert acoustic musical signals into some form of musical notation \cite{benetos2013_Review}, such as piano rolls, sheet music, and Musical Instrument Digital Interface (MIDI), to improve the time-consuming process of manual music transcription. AMT has been applied in automatic annotation of musical information \cite{huang2020_application}, musical education through automatic instrument tutoring \cite{cano2018_application, dunbar2007_application}, and musicological analysis \cite{klapuri2007_application}. In addition, a successful AMT system is also useful for the other tasks in MIR, such as beat tracking \cite{vogl2017_task}, chord recognition \cite{wu2018_task}, and performance classification \cite{kim2020_task}.

Piano transcription is a crucial task of AMT, which typically transcribes piano recordings into a series of note events with pitches, onset/offset timings, and velocities. This task is particularly challenging due to its inherent polyphonic nature, i.e., the multiple pitches are usually in the same frame, thereby causing a complex interaction and overlap of harmonics.  To address this issue, previous methods, taking into account the piano’s acoustic property that the note energy decays after an onset, have mainly focused on adapting models to notes with varying amplitude and harmonics. Inspired by image classification tasks \cite{szegedy2016_CNNimage}, convolutional neural network (CNN)-based methods \cite{kelz2016_CNN, kelz2019_CNN} have been proposed to regress harmonic structures as acoustic representations for audio. On the other hand, recurrent neural network (RNN) methods, such as long short-term memory (LSTM) \cite{bock2012_LSTM} and gated recurrent unit (GRU) \cite{roman2018_GRU}, have been applied to capture the medium- and long-range dependencies between notes. Recently, drawing on advancements in speech recognition tasks, convolutional recurrent neural network (CRNN), which combines a CNN acoustic model with an RNN-based sequential model, has become popular for polyphonic piano transcription. Particularly, Kong et al. \cite{kong2021high} proposed a high-resolution piano transcription system by regressing precise onset and offset times of notes at arbitrary time resolution, achieving effective performance in piano transcription.

However, there are several limitations of the high-resolution system \cite{kong2021high}. First, the harmonics of notes are usually recognized as false positive notes. A potential reason is that the frequency components are calculated with the short-time Fourier transform (STFT), whereas the frequencies that have been chosen to make up the music scale are geometrically spaced, thus yielding components that do not map efficiently to musical frequencies. The second limitation is excessive time and resource consumption due to the large model size. This motivates us to build lightweight architectures for getting around resource constraints and achieve the higher accuracy of transcription.

With this in mind, this work applies an alternative front-end called the Constant-Q Transform (CQT) \cite{brown1991_CQT} instead of the STFT to achieve better simulation of the frequencies in music signals. Moreover, we have designed two architectures: the first is based on a CRNN with dilated convolution to well capture a harmonic structure of music signals on CQT feature, and the second is an encoder-decoder architecture that integrates CRNN with a non-autoregressive Transformer (NR-Transformer) decoder. Compared to the high-resolution system, we show that our systems effectively achieve consistent improvement in note-level metrics.  Specifically, the proposed architectures use significantly fewer parameters of 2.7 million and 0.9 million, respectively, while the high-resolution system uses 20 million parameters. This demonstrates that our systems can achieve the ideal transcription performance without excessive resource consumption.

\section{Related work}
Since the high-resolution system can be seen as the step stone of our work, we here give a detail overview of the high-resolution system in this section.

\begin{figure}[t]
\begin{center}
\includegraphics[width=88mm]{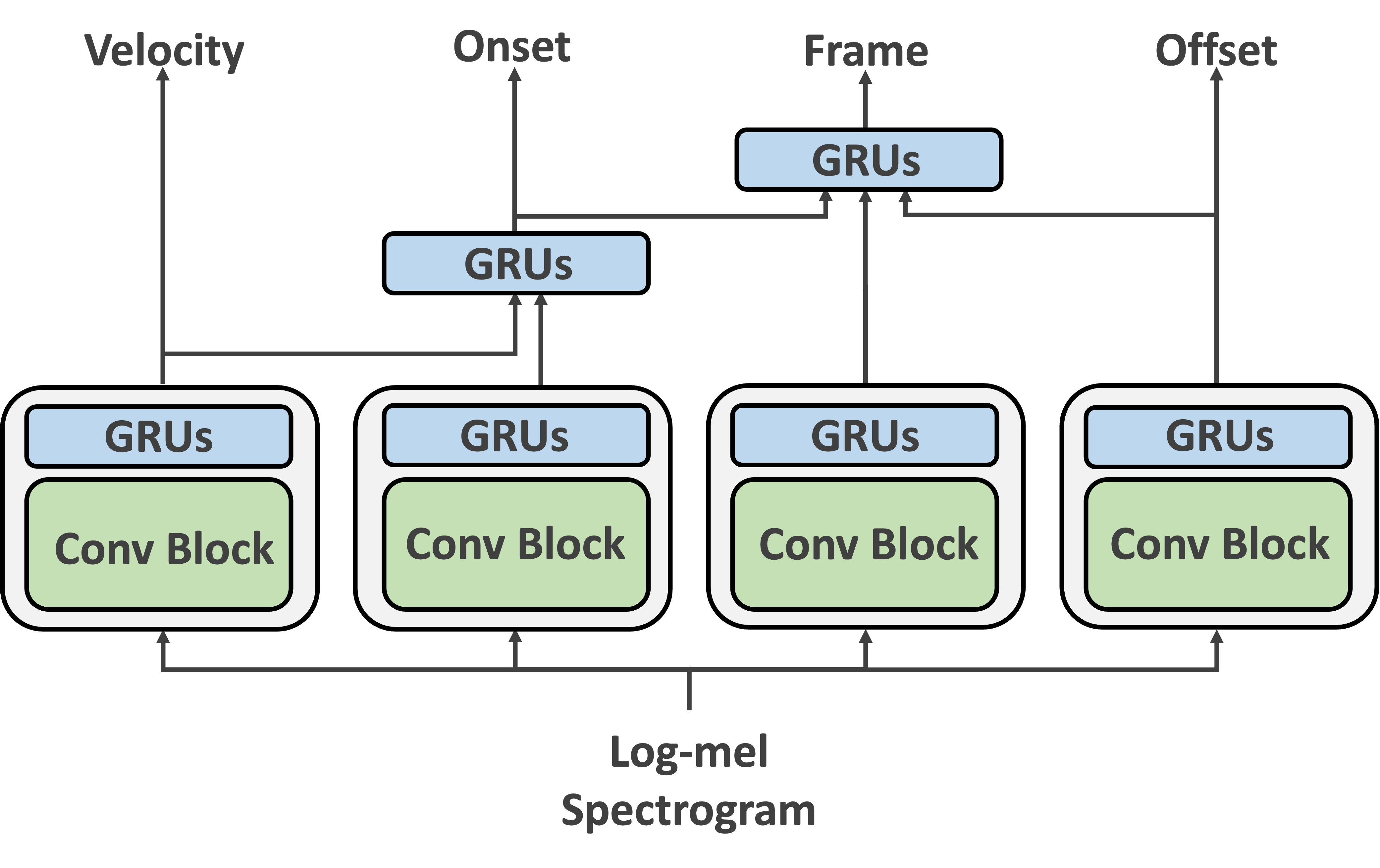}
%\includegraphics[width=70mm]{Fig1-TimesNewRomanPSMT.eps}
%\fbox{ \rule[-17mm]{0pt}{30mm} FIGURE }
\end{center}
\vspace*{-8pt}
\caption{The high-resolution model.}
\label{fig:HRsystem}
\vspace*{-3pt}
\end{figure}

\subsection{Onset and offset times detection}
\label{ssec:onset_offset_detect}
Onsets and offsets represent the beginning and ending of a piano note event, carrying rich information about the piano notes. \cite{Hawthorne2017_Onset&Frame} proposed a dual-objective system for onsets and offsets that conditions the detection of onsets to predict frame-wise outputs. For each piano note, \cite{Hawthorne2017_Onset&Frame} labeled only several consecutive frames of an onset or offset as 1, while other frames was labeled as 0, thereby limiting the transcription resolution. When the precise onset or offset time shifts within a frame, the information about these changes is lost after quantization. Additionally, this labeling method cannot handle cases where the precise onset or offset time is on the boundary between two frames. In light of this, \cite{kong2021high} proposed a new labeling method to address the issue that transcription resolutions are limited by the hop size between adjacent frames, by predicting the continuous onset and offset times of piano notes. A new training target was applied to represent the time difference between the center of a frame and its nearest onset or offset times of a note. The process of encoding the time difference into the training targets can be formulated as
\setlength{\jot}{4pt}
\begin{gather}
\left\{\begin{array}{c}
g\left(\Delta_i\right)=1-\frac{\left|\Delta_i\right|}{J \Delta},|i| \leq J \\
g\left(\Delta_i\right)=0,|i|>J,
\end{array}\right.
\end{gather}
where $\Delta$ and $\Delta_i$ are the frame hop size time and the time difference, respectively. $i$ is the index of a frame, where negative and positive $i$ values indicate the previous and future frame indexes of an onset or offset. $J$ is a hyperparameter that is used to control the sharpness of the targets, i.e., the smaller the $J$, the sharper the targets $g(\Delta_i)$.

\subsection{High-resolution piano transcription system}
Fig.~\ref{fig:HRsystem} shows the architecture of the high-resolution model that we use as the baseline. The log-mel spectrogram with a shape of $T \times F$ calculated from STFT spectrum is the input feature, where $T$ is the number of frames, and $F$ is the number of mel frequency bins. The frame, onset, offset, and velocity tasks share the same acoustic module. This acoustic module, consisting of a convolution block with several convolutional layers and a bidirectional gated recurrent unit (biGRU) layer, is able to extract spectral and temporal information from the log-mel spectrogram. Then, a fully connected layer is applied to output the results of the acoustic model with a shape of $T \times K$, where $K$ is the number of pitch classes. The prediction outputs of velocities are concatenated with outputs of the onsets from the acoustic model. This concatenated data is then fed into a biGRU layer to calculate the final onset predictions. Similarly, the predicted onsets and offsets are used as conditional information to predict frame-wise outputs in the same way.

\begin{figure*}[t]
\begin{center}
\includegraphics[width=180mm]{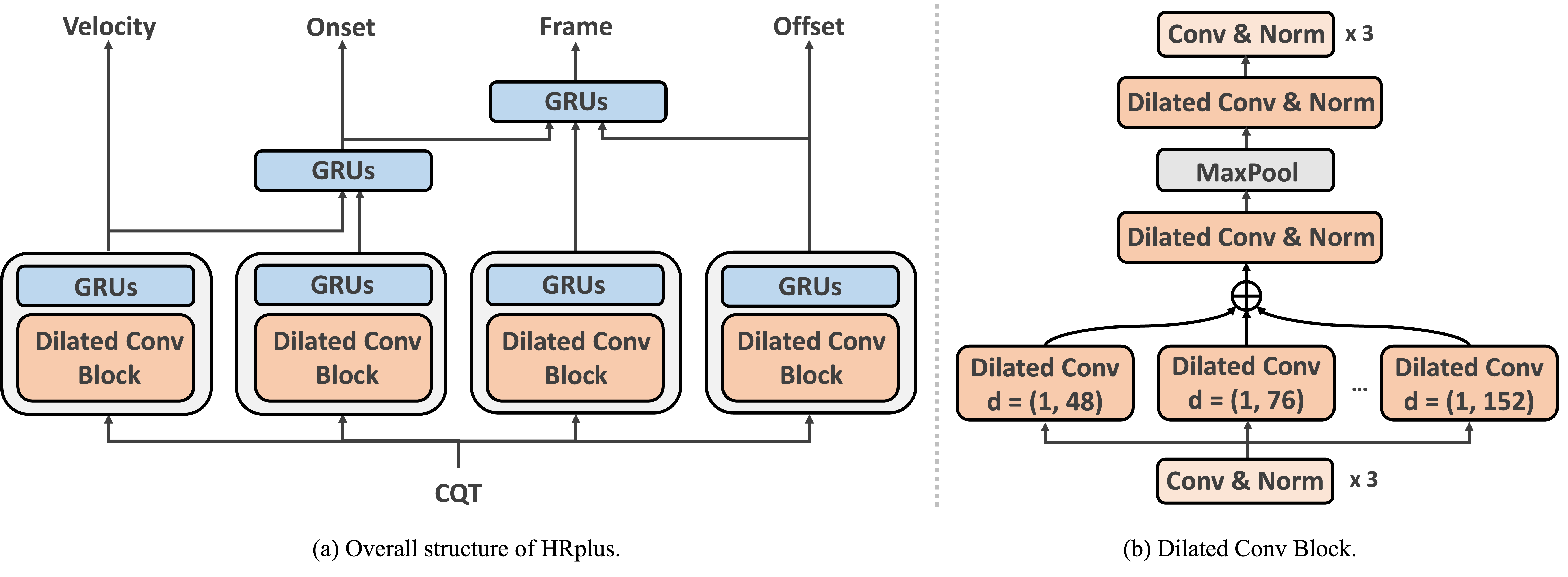}
%\includegraphics[width=70mm]{Fig1-TimesNewRomanPSMT.eps}
%\fbox{ \rule[-17mm]{0pt}{30mm} FIGURE }
\end{center}
\vspace*{-8pt}
\caption{Illustration of HRplus model architecture. Norm denotes instance normalization with a ReLu activation, $d$ denotes the dilation rate.}
\label{fig:HRplus}
\vspace*{-3pt}
\end{figure*}

\begin{figure}[t]
\begin{center}
\includegraphics[width=78mm]{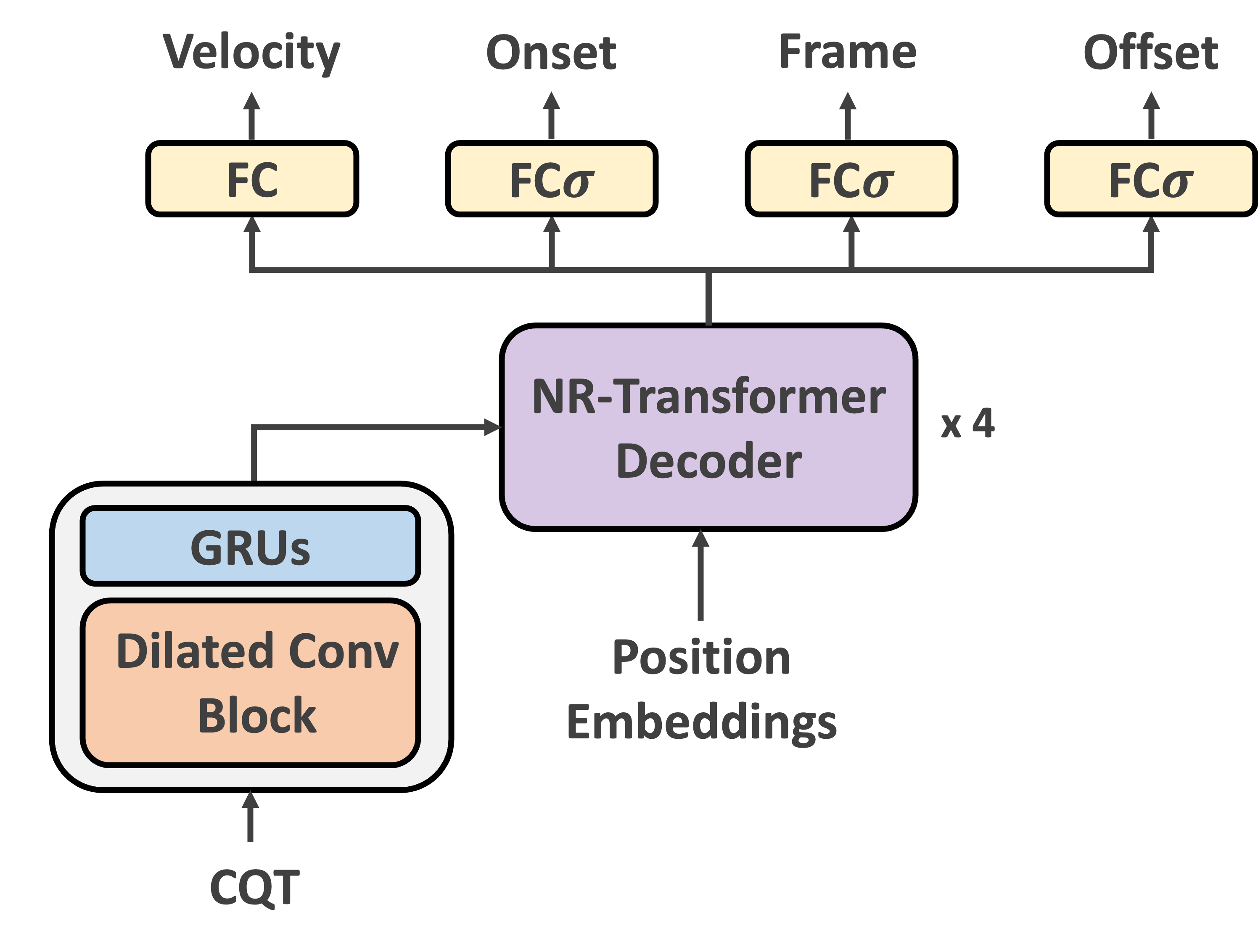}
%\includegraphics[width=70mm]{Fig1-TimesNewRomanPSMT.eps}
%\fbox{ \rule[-17mm]{0pt}{30mm} FIGURE }
\end{center}
\vspace*{-8pt}
\caption{Illustration of HRplus-hybrid model architecture. FC denotes a fully connected layer, $\sigma$ denotes a sigmoid function.}
\label{fig:HRplus-hybrid}
\vspace*{-8pt}
\end{figure}

\section{Proposed method}
At the core of our approach is the idea of effectively simulating the frequencies of music signals to improve the performance of the high-resolution system. We have designed two architectures: the first is based on a CRNN with dilated convolution (See Section~\ref{ssec:CRNN model}), while the second combines a CRNN with a NR-Transformer decoder in an encoder-decoder architecture (See Section~\ref{ssec:hybrid model}).
\subsection{Improved CRNN Model for the High-Resolution System}
\label{ssec:CRNN model}
Fig.~\ref{fig:HRplus}(a) shows the overall architecture of our proposed CRNN model for improved high-resolution system. The CQT spectrogram is applied as input instead of the log-mel spectrogram. Onset, offset, frame and velocity tasks share the same acoustic model that includes a dilated convolution block and a biGRU layer. The output of the velocity serves as conditional information for the onset, while the outputs of both onset and offset condition the frame. We denote this model as HRplus.
\subsubsection{Inputs and Outputs}
\label{ssec:inputs and outputs}
To better simulate the frequencies in music signals, we use CQT as the input representation. Unlike the traditional Fourier transform or the STFT which employ a linear frequency scale, the CQT utilizes a logarithmic frequency scale that closely approximates musical frequencies. As described in \cite{wei2022_harmof0}, the distance $d_k$ between the fundamental frequency $f_0$ and the $k$-th overtone is given by 
\begin{gather}
\begin{aligned}
d_k & =\log _{2^{1 / Q}}\left(k \cdot f_0\right)-\log _{2^{1 / Q}}\left(f_0\right) \\
& =Q \cdot \log _2(k),
\end{aligned}
\end{gather}
where $Q$ is the constant factor of the filter banks that indicates the number of frequency bins per octave.

The outputs of models consist of frame, onset, offset, and velocity events, where onsets and offsets are represented as continuous events, as described in Section~\ref{ssec:onset_offset_detect}. Then, we post-process note-wise events into a set that contains onset time, offset time, and velocities, using the algorithm of the high-resolution system. This post-process proceeds in the following steps:
{\par\noindent\textit{Step 1. Note onset detection.} If the value of an onset event in a frame exceeds the onset threshold, and this value is a local maximum, this frame is detected to contain an onset. Then, the precise onset time is calculated. 
{\par\noindent\textit{Step 2. Velocity scaling.} MIDI files use integers between 0 and 127 to represent the velocity of notes. Since we normalize the dynamic range of velocities from $[0, 127]$ to $[0, 1]$ during model training, the predicted velocities are scaled back to the $[0, 127]$ range if an onset is detected in Step 1.
{\par\noindent\textit{Step 3. Note offset detection.} For the onset detected in Step 1, an offset is detected if the offset prediction output exceeds the offset threshold, or if any frame prediction outputs fall below the frame threshold. Then, the precise offset time is calculated. Additionally, when consecutive onsets of the same pitch are detected, the previous onsets are truncated by adding offsets.

\subsubsection{Acoustic Model for CQT Input Representation}
Considering the challenge of capturing and analyzing multi-scale frequency information scattered in the CQT, we use a dilated convolution block and a biGRU layer as the acoustic model, inspired by HPPNet-sp \cite{wei2022hppnet}. The detail of dilated convolution block is shown in Fig.~\ref{fig:HRplus}(b). The first three convolution layers, each with a kernel size of $7\times7$ and equipped with instance normalization and ReLU activation, extract local information from the CQT input. Subsequently, eight dilated convolution layers with kernel sizes of $1\times3$ apply different dilation rates of 48, 76, 96, 111, 124, 135, 144, and 152, respectively, to capture the harmonic series. These dilation rates, calculated by Eq. (2) with $Q=48$, correspond to the intervals between adjacent harmonics in a harmonic series, i.e., $d_2$, $d_3$, ..., $d_9$ in Eq. (2). The outputs of the eight dilated convolution layers are combined and then fed into a dilated convolution layer with a dilation rate of 48 and a kernel size of $1\times3$, equipped with instance normalization and ReLU activation. Subsequently, a max-pooling layer with a pooling size of 4 along the frequency axis reduces the frequency bins to match the number of pitch classes. Finally, a dilated convolution layer with a kernel size of $1\times3$, a dilation rate of 12, instance normalization, and ReLU activation is applied, followed by three convolution layers with kernel sizes of $5\times1$.

\subsection{Hybrid CRNN-Transformer Encoder-Decoder Model}
\label{ssec:hybrid model}
We have designed an encoder-decoder architecture integrating the acoustic model of HRplus with the NR-Transformer decoder, as shown in Fig.~\ref{fig:HRplus-hybrid}. We denote this model as HRplus-hybrid. The motivations for designing this encoder-decoder architecture are derived from two main considerations: 1) the acoustic model of HRplus could serve exclusively as the encoder of HRplus-hybrid, so as to focus on extracting features from the input and transforming them into a intermediate high-dimensional representation; and 2) the strength of Transformer decoder in capturing long-term dependencies would be leveraged, thanks to which these intermediate representations could be effectively decoded to more accurate outputs even if reducing the model size. The decoder employs the NR-Transformer decoder, which includes four decoder blocks stacked in series. Each NR-Transformer decoder block includes a self-attention module, a cross-attention module, and a feed-forward module. The outputs of the encoder and a trainable positional embedding are used to calculate the cross-attention. Finally, a fully connected layer outputs the predictions for onset, offset, frame, and velocity.

\begin{table*}[t]\small
% \begin{center}
\caption{Transcription results evaluated on the MAESTRO v2 dataset \\ (P: precision, R: recall, \textbf{bold}: best score)}
\label{tab:results_maestrov2}
\centering
\begin{tabular}{ccccclccclccc}
\toprule[1pt]
\multirow{2}{*}{\textbf{Model}} &
  \multirow{2}{*}{\textbf{Params}} &
  \multicolumn{3}{c}{\textbf{Onset}} &
   &
  \multicolumn{3}{c}{\textbf{Onset \& Offset}} &
   &
  \multicolumn{3}{l}{\textbf{Onset, Offset \& Velocity}} \\ \cline{3-5} \cline{7-9} \cline{11-13} \addlinespace[2.5pt]
                                 &   & P (\%) & R (\%) & F1 (\%) &  & P (\%) & R (\%) & F1 (\%) &  & P (\%) & R (\%) & F1 (\%) \\ \midrule
High-resolution \cite{kong2021high} & 20M & 98.17      & 95.35      & 96.72       &  & 83.68      & 81.32      & 
82.47       &  & 82.10      & 79.80      & 80.92       \\
HRplus                    & \textbf{2.7M} & \textbf{98.96}      & \textbf{95.81}      & \textbf{97.34}       &  & \textbf{86.05}      & \textbf{83.36}      & \textbf{84.67}      &  & \textbf{84.21}      & \textbf{81.59}      & \textbf{82.86} \\
\bottomrule[1pt]
\end{tabular}
% \end{center}
\end{table*}

\begin{table*}[t]\small
% \begin{center}
\caption{Transcription results evaluated on the MAESTRO v3 dataset \\ (P: precision, R: recall, \textbf{bold}: best score, \underline{underline}: second best score)}
\label{tab:results_maestrov3}
\centering
\begin{tabular}{ccccclccclccc}
\toprule[1pt]
\multirow{2}{*}{\textbf{Model}} &
  \multirow{2}{*}{\textbf{Params}} &
  \multicolumn{3}{c}{\textbf{Onset}} &
   &
  \multicolumn{3}{c}{\textbf{Onset \& Offset}} &
   &
  \multicolumn{3}{l}{\textbf{Onset, Offset \& Velocity}} \\ \cline{3-5} \cline{7-9} \cline{11-13} \addlinespace[2.5pt]
                                 &   & P (\%) & R (\%) & F1 (\%) &  & P (\%) & R (\%) & F1 (\%) &  & P (\%) & R (\%) & F1 (\%) \\ \midrule
High-resolution {[}reproduced{]} & 20M & 98.22      & 95.26      & 96.69       &  & 83.33      & 80.86      & 82.06       &  & 81.68      & 79.28      & 80.44       \\
HPPNet-sp \cite{wei2022hppnet}                        & \underline{1.2M} & 98.45      & \textbf{95.95}      & \underline{97.18}       &  & 84.88      & 82.76      & 83.80       &  & 83.29      & \underline{81.24}      & 82.24       \\
HRplus                    & 2.7M & \textbf{99.01}      & \underline{95.86}      & \textbf{97.39}       &  & \textbf{86.14}      & \textbf{83.46}      & \textbf{84.76}       &  & \textbf{84.31}      & \textbf{81.69}      & \textbf{82.96}       \\
HRplus-hybrid                  & \textbf{0.9M} & \underline{98.68}      & 94.92      & 96.73       &  & \underline{85.98}      & \underline{82.77}      & \underline{84.32}       &  & \underline{83.91}      & 80.79      & \underline{82.30}       \\ \bottomrule[1pt]
\end{tabular}
% \end{center}
\end{table*}

\subsection{Loss functions}
We use a similar loss calculation as the high-resolution system for the proposed models, with the total loss $L_{total}$ being the sum of the losses for onset $L_{on}$, offset $L_{off}$, frame $L_{fr}$, and velocity $L_{vel}$ as
\begin{gather}
  {L_{total}} = L_{on} + L_{off} + L_{fr} + L_{vel}.
\end{gather}

We denote the continuous target and prediction of onset and offset as $g_{on}$, $\hat{g}_{on}$, $g_{off}$, and $\hat{g}_{off}$, respectively, where they are regarded as the probability of a binary variable (i.e., from 0 to 1), and the binarized target and prediction of frame as $b_{fr}$ and $\hat{b}_{fr}$ (i.e., 0 or 1 for the target and from 0 to 1 for the prediction). The onset loss $L_{on}$, offset loss $L_{off}$, and frame loss $L_{fr}$ are represented as
\begin{gather}
  {L_{on}} = \sum_{t=1}^T \sum_{k=1}^K {L}_{bce}(g_{on}(t,k), \hat{g}_{on}(t,k)), \\
  {L_{off}} = \sum_{t=1}^T \sum_{k=1}^K {L}_{bce}(g_{off}(t,k), \hat{g}_{off}(t,k)), \\
  {L_{fr}} = \sum_{t=1}^T \sum_{k=1}^K {L}_{bce}(b_{fr}(t,k), \hat{b}_{fr}(t,k)).
\end{gather}
$L_{bce}$ is a binary cross-entropy loss function defined as
\begin{gather}
  {L_{bce}(y, \hat{y})} = -y\cdot log(\hat{y})-(1-y)\cdot log(1-\hat{y}),
\end{gather}
where $y$ denotes target and $\hat{y}$ denotes prediction.
Since we predict the velocity only where the onset is detected, the velocity loss $L_{vel}$ is represented as
\begin{gather}
{L_{vel}} = \sum_{t=1}^T \sum_{k=1}^K {b}_{on}(t,k)\cdot{L}_{bce}(b_{vel}(t,k), \hat{b}_{vel}(t,k)),
\end{gather}
where $b_{vel}$ and $\hat{b}_{vel}$ are the binarized target and prediction of velocity, ${b}_{on}\in\{0,1\}^{T\times K}$ indicates the presence or absence of note onsets.

\section{Experimental Evaluations}
\subsection{Datasets}
To evaluate the performance of our systems on the piano transcription task, we used the MIDI and Audio Edited for Synchronous Tracks and Organization (MAESTRO) \cite{hawthorne2018_maestro} dataset, a large-scale piano dataset containing about 200 hours of paired CD-quality audio recordings and MIDI files from ten years of the International Piano-e-Competition. These audio recordings and MIDI files are aligned with around 3 ms accuracy and sliced into individual musical pieces, which are annotated with composer, title, and year of performance. Virtuoso pianists performed on Yamaha Disklaviers, concert-quality acoustic grand pianos, integrated with a high-precision MIDI capture and playback system. To compare against the baselines, we trained and evaluated on MAESTRO v2 and MAESTRO v3 datasets, respectively. We used the train/validation/test split following by the official configuration of MAESTRO dataset without any extensions or augmentations. The total duration of each split in hours are 161.3/19.4/20.5 in MAESTRO v2 and 159.2/19.4/20.0 in MAESTRO v3, respectively.

\subsection{Experimental Setup}
We used PyTorch \cite{paszke2019_pytorch} to implement our systems. The audio recordings were split into 20-second pieces and resampled to 16 kHz so that all the frequencies of the piano could be covered. Then, we down-mixed the audio into a single channel and converted it to a CQT using the nnAudio toolkit \cite{cheuk2020_nnaudio} with a hop length of 320 points, 48 bins per octave, resulting in a total of 352 frequency bins. For high-resolution label for onset and offset, we set the hyperparameter $J = 5$, i.e., each onset or offset affects the regression values of $2 \times J = 10$ frames. 

All the models we proposed were trained with the Adam optimizer \cite{kingma2014_adam} with a batch size of $2$ and a learning rate of $6 \times 10^{-4}$. We used the ReduceLROnPlateau scheduler in PyTorch for learning rate scheduling, employing its default parameters. The best models were determined by the performance on the validation set. At inference, the velocity threshold was set to $0$, while all other thresholds were set to $0.4$. The outputs were converted to MIDI events as described in Sections~\ref{ssec:inputs and outputs}.

\subsection{Baselines}
Since the proposed models are optimized based on the high-resolution system, and the acoustic model is inspired by HPPNet-sp, we use both High-resolution \cite{kong2021high} and HPPNet-sp \cite{wei2022hppnet} as baselines. The results are shown in Table~\ref{tab:results_maestrov2} and Table~\ref{tab:results_maestrov3}. High-resolution applies the high-resolution labels for note onset and offset to construct the piano transcription system. HPPNet-sp uses dilated convolution and frequency-grouped LSTM to model the harmonic structure and pitch-invariance over time in piano transcription.

\subsection{Evaluation Metrics}
For evaluation metrics of the piano transcription systems, we used note-level metrics involving the standard precision, recall, and F1 score. These metrics match each predicted note with a ground truth note, considering onset times, pitches, and optional offset times and velocities. We used the mir\_eval library \cite{raffel2014_mireval} for all metric calculations. Following the default configuration of mir\_eval, we applied a 50 ms tolerance for note onsets, a 20\% offset ratio for note offsets, and a tolerance of 0.1 for velocities.

\subsection{Experimental Results}
Table~\ref{tab:results_maestrov2} shows the evaluation results for the MAESTRO v2. For the MAESTRO v2 dataset, HRplus significantly outperforms High-resolution on all metrics. Besides, HRplus uses 2.7 million parameters, markedly fewer than the 20 million parameters used by High-resolution. This demonstrates that our model architecture is more lightweight and cost-efficient.

To provide a more comprehensive comparison between our proposed models and the baselines, we train and evaluate High-resolution on the MAESTRO v3 dataset. The results are shown in Table~\ref{tab:results_maestrov3}. From the baseline perspective, we observe that HPPNet-sp outperforms High-resolution in note-level metrics. Meanwhile, HPPNet-sp also has a significantly smaller model size compared to High-resolution. When comparing our proposed models with the baselines, it is observed that HRplus wins in eight out of nine test metrics, especially in the F1 score, which is significantly better than all other systems. Moreover, we find that even with the smallest model size of only 0.9 million parameters, HRplus-hybrid still achieves the second best results in six out of nine test metrics. It outperforms all the baselines in F1 score for Onset \& Offset and Onset, Offset \& Velocity. Specifically, HRplus-hybrid surpasses the High-resolution system in both F1 score and Precision across all metrics. This demonstrates our proposed models, combining the advantages of High-resolution and HPPNet, are very effective in improving AMT performance as well as reducing resource consumption.

We specifically analyze the model size among different systems. First, since the dilated convolution can significantly reduce the number of parameters, the model size of HRplus is greatly reduced compared to that of High-resolution. Second, in contrast to the approach taken by HRplus which utilizes different GRU+dilated conv block for individual outputs, HRplus-hybrid shares one GRU+dilated conv block across all outputs, thus achieving the smallest model size.

In addition, we note that the results of HR-hybrid are slightly inferior to those of HRplus. A potential reason is that HRplus-hybrid does not utilize conditional information in its prediction outputs. Conversely, the use of conditional information, especially onset information, has significantly enhanced the final transcription performance, as evidenced in the studies of \cite{Hawthorne2017_Onset&Frame, kong2021high, wei2022hppnet}.

\section{Conclusions}
In this paper, we utilize the CQT as the input representation instead of the STFT to better adapt to musical signals, thereby improving the transcription performance of the high-resolution system. We design two architectures: the first is based on a CRNN with dilated convolution, and the second is an encoder-decoder architecture that integrates CRNN with a NR-Transformer decoder. The experimental results demonstrate that our proposed methods can effectively achieve consistent improvements in note-level metrics without any extensions or augmentations. Specifically, the proposed architectures use only 2.7 million and 0.9 million parameters, respectively, compared with the 20 million used by the high-resolution system. Therefore, our systems can achieve the ideal transcription without excessive resource consumption. In the future, we will extend our models to other instruments and introduce transfer learning methods to enhance transcription performance.

\section*{Acknowledgment}
This work was partly supported by JST CREST JPMJCR19A3, Japan. In addition, this work was also financially supported by JST SPRING, Grant Number JPMJSP2125. The author would like to take this opportunity to thank the “THERS Make New Standards Program for the Next Generation Researchers.”

\printbibliography

@article{benetos2018_Review,
  title={Automatic music transcription: An overview},
  author={Benetos, Emmanouil and Dixon, Simon and Duan, Zhiyao and Ewert, Sebastian},
  journal={IEEE Signal Processing Magazine},
  volume={36},
  number={1},
  pages={20--30},
  year={2018},
  publisher={IEEE}
}

@article{benetos2013_Review,
  title={Automatic music transcription: challenges and future directions},
  author={Benetos, Emmanouil and Dixon, Simon and Giannoulis, Dimitrios and Kirchhoff, Holger and Klapuri, Anssi},
  journal={Journal of Intelligent Information Systems},
  volume={41},
  pages={407--434},
  year={2013},
  publisher={Springer}
}

@inproceedings{huang2020_application,
  title={Pop music transformer: Beat-based modeling and generation of expressive pop piano compositions},
  author={Huang, Yu-Siang and Yang, Yi-Hsuan},
  booktitle={Proceedings of the 28th ACM international conference on multimedia},
  pages={1180--1188},
  year={2020}
}

@article{cano2018_application,
  title={Music technology and education},
  author={Cano, Estefan{\'\i}a and Dittmar, Christian and Abe{\ss}er, Jakob and Kehling, Christian and Grollmisch, Sascha},
  journal={Springer Handbook of Systematic Musicology},
  pages={855--871},
  year={2018},
  publisher={Springer}
}

@inproceedings{dunbar2007_application,
  title={Music Transcription as Pedagogy: Discussion of a Cross-disciplinary Approach to Teacher Preparation},
  author={Dunbar-Hall, Peter},
  booktitle={Celebrating Musical Communities: Proceedings of the 40th Anniversary National Conference, Perth 6th-8th July 2007},
  pages={89--93},
  year={2007},
  organization={Australian Society for Music Education Nedlands, WA}
}

@article{klapuri2007_application,
  title={Signal processing methods for music transcription},
  author={Klapuri, Anssi and Davy, Manuel},
  year={2007},
  publisher={Springer Science \& Business Media}
}

@article{wu2018_task,
  title={Automatic audio chord recognition with MIDI-trained deep feature and BLSTM-CRF sequence decoding model},
  author={Wu, Yiming and Li, Wei},
  journal={IEEE/ACM Transactions on Audio, Speech, and Language Processing},
  volume={27},
  number={2},
  pages={355--366},
  year={2018},
  publisher={IEEE}
}

@inproceedings{vogl2017_task,
  title={Drum Transcription via Joint Beat and Drum Modeling Using Convolutional Recurrent Neural Networks.},
  author={Vogl, Richard and Dorfer, Matthias and Widmer, Gerhard and Knees, Peter},
  booktitle={Proceedings of the International Society for Music Information Retrieval (ISMIR)},
  pages={150--157},
  year={2017}
}

@article{kim2020_task,
  title={Deep composer classification using symbolic representation},
  author={Kim, Sunghyeon and Lee, Hyeyoon and Park, Sunjong and Lee, Jinho and Choi, Keunwoo},
  journal={arXiv preprint arXiv:2010.00823},
  year={2020}
}

@inproceedings{szegedy2016_CNNimage,
  title={Rethinking the inception architecture for computer vision},
  author={Szegedy, Christian and Vanhoucke, Vincent and Ioffe, Sergey and Shlens, Jon and Wojna, Zbigniew},
  booktitle={Proceedings of the IEEE conference on computer vision and pattern recognition},
  pages={2818--2826},
  year={2016}
}

@article{kelz2016_CNN,
  title={On the potential of simple framewise approaches to piano transcription},
  author={Kelz, Rainer and Dorfer, Matthias and Korzeniowski, Filip and B{\"o}ck, Sebastian and Arzt, Andreas and Widmer, Gerhard},
  journal={arXiv preprint arXiv:1612.05153},
  year={2016}
}

@inproceedings{kelz2019_CNN,
  title={Multitask learning for polyphonic piano transcription, a case study},
  author={Kelz, Rainer and B{\"o}ck, Sebastian and Widnaer, Cierhard},
  booktitle={2019 International Workshop on Multilayer Music Representation and Processing (MMRP)},
  pages={85--91},
  year={2019},
  organization={IEEE}
}

@inproceedings{bock2012_LSTM,
  title={Polyphonic piano note transcription with recurrent neural networks},
  author={B{\"o}ck, Sebastian and Schedl, Markus},
  booktitle={2012 IEEE international conference on acoustics, speech and signal processing (ICASSP)},
  pages={121--124},
  year={2012},
  organization={IEEE}
}

@inproceedings{roman2018_GRU,
  title={An End-to-end Framework for Audio-to-Score Music Transcription on Monophonic Excerpts},
  author={Rom{\'a}n, Miguel A and Pertusa, Antonio and Calvo-Zaragoza, Jorge},
  booktitle={Proceedings of the International Society for Music Information Retrieval (ISMIR)},
  pages={34--41},
  year={2018}
}

@article{brown1991_CQT,
  title={Calculation of a constant Q spectral transform},
  author={Brown, Judith C},
  journal={The Journal of the Acoustical Society of America},
  volume={89},
  number={1},
  pages={425--434},
  year={1991},
  publisher={Acoustical Society of America}
}

@article{kong2021high,
  title={High-resolution piano transcription with pedals by regressing onset and offset times},
  author={Kong, Qiuqiang and Li, Bochen and Song, Xuchen and Wan, Yuan and Wang, Yuxuan},
  journal={IEEE/ACM Transactions on Audio, Speech, and Language Processing},
  volume={29},
  pages={3707--3717},
  year={2021},
  publisher={IEEE}
}

@inproceedings{Hawthorne2017_Onset&Frame,
  title={Onsets and Frames: Dual-Objective Piano Transcription},
  author={Curtis Hawthorne and Erich Elsen and Jialin Song and Adam Roberts and Ian Simon and Colin Raffel and Jesse Engel and Sageev Oore and Douglas Eck},
  booktitle={Proceedings of the International Society for Music Information Retrieval (ISMIR)},
  pages={50--57},
  year={2018}
}

@inproceedings{wei2022_harmof0,
  title={Harmof0: Logarithmic scale dilated convolution for pitch estimation},
  author={Wei, Weixing and Li, Peilin and Yu, Yi and Li, Wei},
  booktitle={2022 IEEE International Conference on Multimedia and Expo (ICME)},
  pages={1--6},
  year={2022},
  organization={IEEE}
}

@inproceedings{wei2022hppnet,
  title={Hppnet: Modeling the harmonic structure and pitch invariance in piano transcription},
  author={Wei, Weixing and Li, Peilin and Yu, Yi and Li, Wei},
  booktitle={Proceedings of the International Society for Music Information Retrieval (ISMIR)},
  year={2022}
}

@inproceedings{
hawthorne2018_maestro,
title={Enabling Factorized Piano Music Modeling and Generation with the {MAESTRO} Dataset},
author={Curtis Hawthorne and Andriy Stasyuk and Adam Roberts and Ian Simon and Cheng-Zhi Anna Huang and Sander Dieleman and Erich Elsen and Jesse Engel and Douglas Eck},
booktitle={International Conference on Learning Representations},
year={2019},
url={https://openreview.net/forum?id=r1lYRjC9F7},
}

@article{paszke2019_pytorch,
  title={Pytorch: An imperative style, high-performance deep learning library},
  author={Paszke, Adam and Gross, Sam and Massa, Francisco and Lerer, Adam and Bradbury, James and Chanan, Gregory and Killeen, Trevor and Lin, Zeming and Gimelshein, Natalia and Antiga, Luca and others},
  journal={Advances in neural information processing systems},
  volume={32},
  year={2019}
}

@article{cheuk2020_nnaudio,
  title={nnaudio: An on-the-fly gpu audio to spectrogram conversion toolbox using 1d convolutional neural networks},
  author={Cheuk, Kin Wai and Anderson, Hans and Agres, Kat and Herremans, Dorien},
  journal={IEEE Access},
  volume={8},
  pages={161981--162003},
  year={2020},
  publisher={IEEE}
}

@article{kingma2014_adam,
  title={Adam: A method for stochastic optimization},
  author={Kingma, Diederik P and Ba, Jimmy},
  journal={arXiv preprint arXiv:1412.6980},
  year={2014}
}

@inproceedings{raffel2014_mireval,
  title={MIR\_EVAL: A Transparent Implementation of Common MIR Metrics.},
  author={Raffel, Colin and McFee, Brian and Humphrey, Eric J and Salamon, Justin and Nieto, Oriol and Liang, Dawen and Ellis, Daniel PW and Raffel, C Colin},
  booktitle={Proceedings of the International Society for Music Information Retrieval (ISMIR)},
  volume={10},
  pages={2014},
  year={2014}
}

\end{document}